\documentstyle[prb,aps,amsfonts,amssymb,floats,epsfig]{revtex}
\begin{document}
\draft
\twocolumn[\hsize\textwidth\columnwidth\hsize\csname
@twocolumnfalse\endcsname
\title{Surface electronic structure of  Sr$_{2}$RuO$_{4}$}
\author{K.M. Shen, A. Damascelli, D.H. Lu, N.P. Armitage, F. Ronning, D.L. Feng, C. Kim, and Z.-X. Shen}
\address{Department of Applied Physics, Physics, and Stanford Synchrotron Radiation Laboratory
\\Stanford University, Stanford, California 94305}
\author{D.J. Singh and I.I. Mazin}
\address{Naval Research Laboratory, Code 6391, Washington, D.C. 20375}
\author{S. Nakatsuji, Z.Q. Mao, and Y. Maeno}
\address{Department of Physics, Kyoto University, Kyoto 606-8502, and CREST-JST, Kawagushi,
Saitama 332-0012, Japan}
\author{T. Kimura and Y. Tokura}
\address{Department of Applied Physics, University of Tokyo, Tokyo 113-8656,
and JRCAT, Tsukuba, 305-0046, Japan}

%
\maketitle
\begin{abstract}
We have addressed the possibility of surface ferromagnetism in
Sr$_2$RuO$_4$ by investigating its surface electronic states by
angle-resolved photoemission spectroscopy (ARPES). By cleaving
samples under different conditions and using various photon
energies, we have isolated the surface from the bulk states. A
comparison with band structure calculations indicates that the
ARPES data are most readily explained by a \emph{nonmagnetic}
$\sqrt{2} \times \sqrt{2}$ surface reconstruction.
\end{abstract}


\vspace{-0.5cm}

%
\vskip2pc]
%

Following the discovery of superconductivity (SC) at 1 K in the
layered perovskite Sr$_{2}$RuO$_{4}$, \cite{maeno} the exact
nature of its SC pairing mechanism has attracted a great deal of
interest. While it shares the same structure as the archetypal
cuprate parent compound La$_{2}$CuO$_{4}$, RuO$_{2}$ planes
replace the CuO$_{2}$ planes thus resulting in an anisotropic
Fermi liquid \cite{mackenzie} instead of a strongly correlated
charge transfer insulator. Furthermore, there is evidence that
Sr$_{2}$RuO$_{4}$ exhibits spin-triplet pairing with a $p$-wave
order parameter,\cite{ishida} as opposed to the spin-singlet,
$d$-wave symmetry found in the cuprates. Although it is now widely
believed that the unconventional nature of SC in this compound is
mediated by spin fluctuations, the exact nature of this
interaction is still unresolved. Originally, it was suggested that
ferromagnetic (FM) spin fluctuations were responsible for
mediating the SC as inferred from theoretical calculations,
\cite{mazin1} NMR measurements,\cite{imai} and ferromagnetism in
closely related SrRuO$_{3}$. However, more recent evidence has
suggested that this simple picture may be incomplete.
Antiferromagnetism (AFM) in Ca$_{2}$RuO$_{4}$, the observation of
incommensurate peaks at $\mathbf{Q}$ $= (0.6\pi, 0.6\pi, 0)$ by
neutron scattering, \cite{sidis} and calculations which show
strong nesting at $\mathbf{Q}$ $= (2\pi/3, 2\pi/3, 0)$
\cite{mazin2} all seem to imply AFM correlations should not be
neglected, leaving the nature of SC open to speculation.

Recently, an analysis of low-energy electron diffraction data from
Sr$_2$RuO$_4$ indicated that a $\sqrt{2} \times \sqrt{2}$
reconstruction was induced by the freezing of a soft zone boundary
phonon into a static lattice distortion at the surface, and
comparisons with band structure calculations predicted that the
resulting surface was FM. \cite{matzdorf} This conjecture also
appears conceivable in light of speculation based on recent ARPES
results \cite{damascelli} as well as earlier theoretical
calculations.\cite{degroot} If FM does exist on the surface of
Sr$_{2}$RuO$_{4}$, such a result should be indicative of strong
ferromagnetic tendencies in the bulk and thus possibly relevant to
microscopic theories which describe the mechanism of SC. This
speculation becomes even more intriguing in light of recent STM
measurements\cite{morpurgo} which suggest the opening of a
superconducting gap with T$_{c}$ = 1.4 K, perhaps hinting that the
surface layer may be superconducting, and raises the possibility
that the surface of Sr$_2$RuO$_4$ may exhibit the rare coexistence
of SC and FM. However, as this proposed surface FM has never been
confirmed, it becomes imperative to reinvestigate the
\emph{surface} electronic structure to definitively verify or
exclude surface FM.

In this paper, we present a detailed, high-resolution ARPES study
of the surface electronic structure of Sr$_{2}$RuO$_{4}$. While
our earlier work \cite{damascelli} ascertained that the
\emph{bulk} Fermi surface (FS) topology extracted by ARPES was
indeed in excellent agreement with both theory \cite{oguchi,singh}
and de Haas-van Alphen (dHvA) results, \cite{mackenzie} the
precise nature of the surface-derived states, which could be
nonmagnetic (NM) or FM, remained somewhat ambiguous. In
particular, our earlier depiction of the surface electronic
structure failed to explicate the presence of the intense,
surface-derived peak at $(\pi, 0)$, leaving speculation that its
existence could be deemed a manifestation of surface FM. To
clarify this uncertainty, we have performed a comprehensive ARPES
study with various photon energies and polarizations in
conjunction with detailed band structure calculations which now
account for the surface reconstruction. By comparing these
calculations with our ARPES data, we conclude that our results are
consistent with the NM scenario and exhibit no experimentally
detectable trace of surface FM down to 10 K.

\begin{figure}[t!]
\centerline{\epsfig{figure=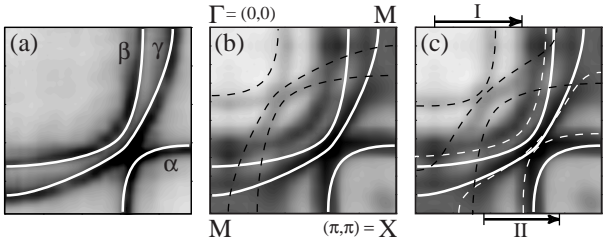,width=8cm,clip=}}
\vspace{0cm} \caption{(a) : E$_F$ intensity map of Sr$_2$RuO$_4$
cleaved at 180 K and measured at 10 K with $h\nu$ = 28 eV and
overlaid theoretical FSes. (b) and (c) show an intensity map from
a sample cleaved and measured at 10 K. (b) shows calculated bulk
FSes (white) with trivial folded FSes (dashed), while (c) shows
calculated 6$^\circ$ NM reconstruction with both primary and
folded surface FSes (dashed white and dashed black).} \label{FS}
\end{figure}

ARPES data was taken at the Stanford Synchrotron Radiation
Laboratory using a Scienta SES-200 analyzer with typical
resolutions of $\Delta$E \mbox{$<$ 13 meV} and \mbox{$\Delta\theta
\approx$ 0.2$^\circ$}. Sr$_{2}$RuO$_{4}$ single crystals were
first aligned by Laue diffraction and then cleaved \emph{in situ}
at a pressure of better than $5 \times 10^{-11}$ torr and at
various temperatures described below. All spectra were taken at 10
K, in both the first and second Brillouin zones; surface features
showed slight enhancement in the second zone.

\begin{figure}[b!]
\centerline{\epsfig{figure=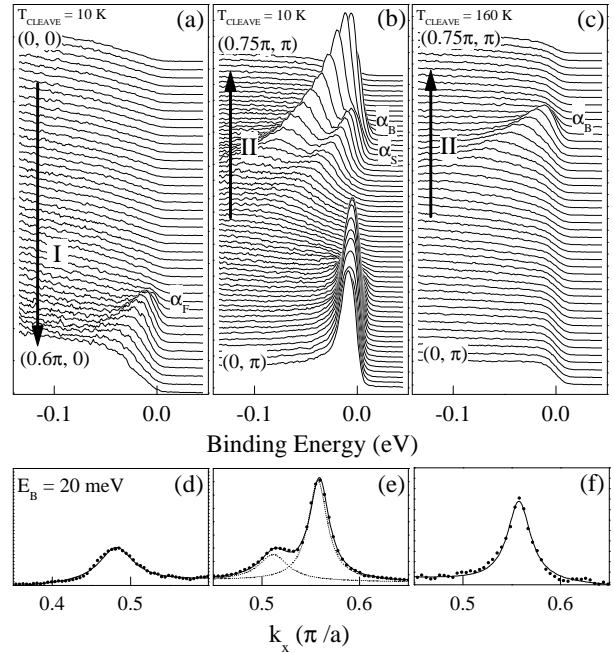,width=8cm,clip=}}
\vspace{0cm} \caption{EDCs in panels (a), (b), and (c) with along
cuts I and II shown in Figure \ref{FS}. Corresponding MDCs at
E$_B$ = 20 meV taken from (a), (b), and (c) are shown in panels
(d), (e) and (f), respectively. Data from (a), (b), (d), and (e)
were taken on a sample cleaved at 10 K, while data from (c) and
(f) were taken on a sample cleaved at 160 K. $\alpha_F$,
$\alpha_S$, and $\alpha_B$ refer to folded, surface, and bulk
$\alpha$ bands, respectively.}\label{EDCs}
\end{figure}

Figure \ref{FS}a shows an E$_F$ intensity map (integrated signal
within E$_F$ $\pm$ 5 meV) of a sample cleaved at 180 K and
measured at 10 K. As discussed in Ref. \onlinecite{damascelli},
cleaving the sample at elevated temperatures preferentially
suppresses the surface intensity; we speculate that the increased
rate of thermally activated oxygen diffusion results in a more
disordered surface layer. The resulting intensity map thereby
primarily reflects the bulk contribution, and the calculated bulk
FSes from Ref. \onlinecite{singh} are overlaid and in excellent
agreement. When cleaving at lower temperatures, the surface states
were well preserved and also apparent in our data, in addition to
the bulk states. This additional surface contribution is clearly
visible in the intensity maps in Figures \ref{FS}b and \ref{FS}c
and somewhat complicates the situation. Our original conjecture,
in Ref. \onlinecite{damascelli} and shown in Figure \ref{FS}b,
accounted for the surface states by considering them to be the
same as those of the bulk, except for a rigid folding due to the
$\sqrt{2} \times \sqrt{2}$ surface reconstruction; the
reconstruction arises from rotations of the RuO$_6$ surface
octahedra which cause a doubling of the surface unit
cell.\cite{matzdorf} Despite the approximate agreement, this
overly simplistic picture fails to explain the origin of the
strong peak (bottom of Figure \ref{EDCs}b) at M, which influenced
earlier ARPES reports to erroneously designate the bulk
$\gamma$-FS as hole-like.\cite{yokoya,lu} This apparent
discrepancy also led us to initially posit that surface FM might
be responsible for this state at M. However, after calculating the
precise effects of the surface distortion on the band structure,
we find that the NM reconstruction alone can potentially drive the
surface $\gamma$-FS hole-like, thus explaining the peak at M; this
more accurate NM scenario is depicted in Figure \ref{FS}c.

Nonetheless, since surface FM might still account for some of the
experimentally observed features, it becomes crucial to examine
the surface states in greater detail. In particular, surface FM
would cause the surface states to split into minority and majority
bands, effectively doubling the number of surface-derived bands.
In order to distinguish between the NM and FM scenarios, we focus
on ARPES spectra taken along lines I and II in Figure \ref{FS}c.
For the NM surface, we would expect to see one band, $\alpha_F$,
crossing along I and two crossings, $\alpha_S$ and $\alpha_B$,
along II. Any additional bands beyond those predicted for the NM
surface would be strong evidence for surface FM, and should be
readily apparent in the ARPES data.

To address this issue, we first focus on spectra taken along I,
shown in Figure \ref{EDCs}a, using 24 eV photons polarized along
the Ru-O bond direction; different photon energies and
polarizations yielded similar results. This region is particularly
suitable for an investigation into potential surface FM since it
is far removed from the bulk electronic states. Examining the
energy distribution curves (EDCs) in Figure \ref{EDCs}a, we see
only a single electronic state from the folded $\alpha$-FS,
denoted as $\alpha_F$, as is expected from the NM scenario shown
in Figure \ref{FS}c. Conversely, additional bands reflecting the
spin splitting would be expected for a FM surface. In Figure
\ref{EDCs}d, we show a momentum distribution curve (MDC) of data
from \ref{EDCs}a, where the photoemitted electron intensity is
displayed as a function of momentum at a fixed binding energy of
E$_B$ = 20 meV and fitted to a single Lorentzian lineshape on a
linear background. By analyzing our data in this fashion, we are
able to track the dispersion of $\alpha_{F}$ yielding
$v_{F}^{F}\!=\!$ \mbox{0.7 eV$\cdot$\AA}. Therefore our
measurements along I yield only a single surface band, consistent
with the nonmagnetic scenario of Figure \ref{FS}c.

To further reinforce this result, we now consider data along II
shown in Figure \ref{EDCs}b, taken under the same conditions as I,
but in the second zone. In both the EDCs and MDCs, one can clearly
observe two distinct peaks. By fitting the MDCs to a
double-Lorentzian form, shown in Figure \ref{EDCs}e, and tracking
their dispersion to E$_F$, one can determine the Fermi velocities
of the two bands. From this analysis, we determine the velocity of
the first band, $\alpha_{B}$, to be $v_{F}^{B}$ = \mbox{1.1
eV$\cdot$\AA}, while for the second, $\alpha_{S}$, $v_{F}^{S}$ =
\mbox{0.7 eV$\cdot$\AA}. On another sample cleaved at 160 K with
the measurement taken in otherwise identical conditions,
$\alpha_{S}$ is suppressed, as shown in Figure \ref{EDCs}c, and
the remaining state is the bulk-derived $\alpha_B$; both
$\alpha_B$ features in \ref{EDCs}b and \ref{EDCs}c have the same
$v_{F}$ and Fermi crossing position. Also note that cleaving at
elevated temperatures completely suppresses the strong peak at the
bottom of \ref{EDCs}b, which is also responsible for the weight at
M in Figures \ref{FS}b and \ref{FS}c. Furthermore, we are able to
conclude that $\alpha_{F}$ is simply the folded counterpart of the
surface-derived $\alpha_{S}$, and not the counterpart of the
bulk-derived $\alpha_B$, since $\alpha_{S}$ and $\alpha_{F}$ have
matching $v_F$ and Fermi crossings in the reduced zone. Therefore,
the absence of additional bands along II is consistent with our
results from I and our conclusion of a NM surface.

\begin{figure}[b!]
\centerline{\epsfig{figure=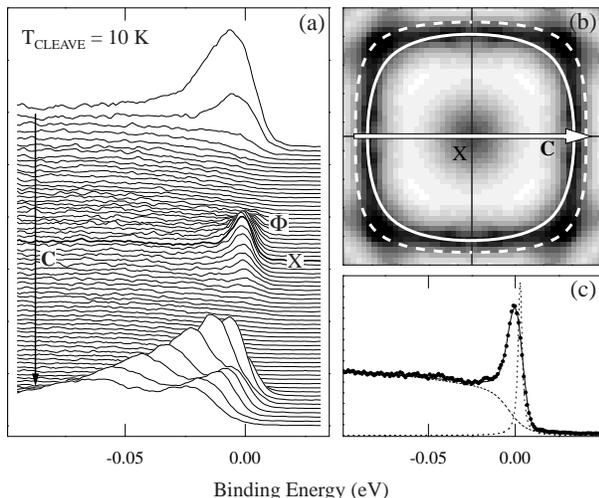,width=8cm,clip=}}
\vspace{0cm} \caption{ARPES data taken near X with $h\nu$ = 24 eV
at 10 K on a sample cleaved 10 K. EDCs in (a) were taken along cut
C. (b) shows an E$_F$ intensity map ($\pm$ 5 meV) around the
region of the bulk and surface $\alpha$-FSes, shown in solid and
dashed white, respectively. (c) shows the EDC from X and the
corresponding fit. The background curve and the Lorentzian used in
fitting are also shown as dotted lines.} \label{Xpoint}
\end{figure}

Examining EDCs taken over the entire zone, virtually all observed
surface states can be well accounted for by considering only a NM
surface. The only unexpected feature was a small peak localized
around X, hereafter denoted as $\Phi$, as shown in Figures
\ref{Xpoint}a and \ref{Xpoint}b, which was most strongly enhanced
at $h\nu$ = 24 eV. Close inspection of the spectrum at X in Figure
\ref{Xpoint}c reveals that the peak position is located at E$_F$
($\pm$ 1 meV) and the leading edge is 6 meV above E$_F$,
indicating that this peak originates from above E$_F$; the peak in
the EDCs is the product of the rising tail of the quasiparticle
peak and the falling edge of the Fermi-Dirac function. This was
confirmed by fitting the data using a background taken from
$(0.8\pi, \pi)$ and a Lorentzian peak, both multiplied by a
Fermi-Dirac function and convolved with our resolution ($\Delta$E
= 8 meV), allowing us to estimate a peak position of 3 meV above
E$_{F}$ and an intrinsic FWHM of 3 meV. Although there may be some
slight inaccuracies in the fitting procedure, all attempts to fit
the spectra using a below-E$_F$ peak proved wholly unsuccessful.
Moreover, since no corresponding band can be seen to disperse from
below E$_F$, we can conclude that $\Phi$ arises from an unoccupied
band whose minimum at X almost grazes E$_F$. Naively, one might
infer that this somewhat unexpected feature could be interpreted
as evidence for surface FM. However, as will be discussed below,
our calculations, in fact, even predict the appearance of $\Phi$,
which arises from the distortion of the surface layer.

In order to gain deeper insight into the effects of the surface
reconstruction, we have performed both NM and FM calculations
similar to those reported in Ref.\onlinecite{singh}, but including
rotations by a fixed angle in all RuO$_2$ planes, resulting in a
monoclinic $C2/m$ symmetry. We will hereafter refer to these
rotated bulk calculations as pertaining to the surface, and this
assumption can be justified because of the extremely 2D nature of
the electronic structure; any effects from the surface termination
should be far weaker than those of the rotations of the octahedra,
and is demonstrated by the excellent agreement of our NM rotated
calculations with the corresponding surface calculations performed
by Fang. \cite{matzdorf,fang}

\begin{figure}[t!]
\centerline{\epsfig{figure=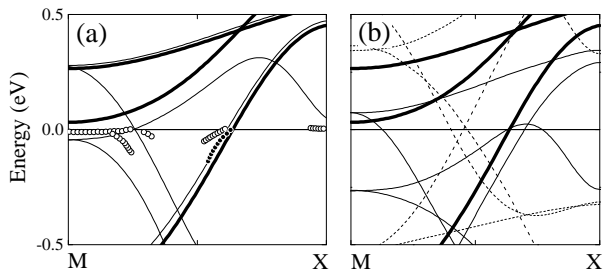,width=8cm,clip=}}
\vspace{0cm} \caption{(a) : Band structure calculations along M-X
for bulk and 6$^\circ$ NM surface (thick and thin lines) along
with ARPES dispersion for bulk and surface states (solid and open
circles). (b) : Bulk, minority surface, and majority surface bands
in thick, thin, and dashed lines for a 6$^\circ$ FM surface.}
\label{bands}
\end{figure}

First, we consider the results from the NM calculations; Figure
\ref{bands}a shows the results of calculations along the
tetragonal M-X direction for rotations of $\theta$ = 0$^\circ$
(bulk) and $\theta$ = 6$^\circ$ (NM surface). An angle of
6$^\circ$ was chosen since it is within the error bars of the
structural data \cite{matzdorf} and also gives good agreement with
our ARPES data, especially the placement of $\Phi$. The NM surface
and bulk bands are shown in Figure \ref{bands}a, and the
experimentally determined values are overlaid and in good
agreement with theory. \cite{velocityratio} We also note that our
estimate of the quasiparticle renormalization $v_{F}^{calc}$ /
$v_{F}^{B}$ = 3.2 for the bulk $\alpha$-FS along M-X is in
excellent agreement with $m^\ast / m_{band}$ = 3.3 from dHvA
\cite{mackenzie}, and also close to theoretical estimates of
$\approx$ 2.5.\cite{liebsch} Calculations for $\theta$ = 6$^\circ$
produce the dashed FSes in Figure \ref{FS}c and the rotation
induces various effects.

First, the extended van Hove singularity (evHs) at M, which is 60
meV above E$_F$ in the bulk calculations, is pushed 40 meV below
E$_F$ due to the repulsion between the $d_{xy}$ and $d_{3z^2 -
r^2}$ bands, which is allowed only in the lowered symmetry of the
distorted surface. This results in the topology of the surface
$\gamma$-FS changed from electron-like to hole-like, as shown in
Figure \ref{FS}c, and is also consistent with independent
calculations from Ref. \onlinecite{liebsch2}. Furthermore, the
dispersion of this feature, in agreement with
Refs.\onlinecite{yokoya} and \onlinecite{lu}, is consistent with
the saddlepoint topology predicted by theory.

Secondly, the lower symmetry on the surface also allows for
hybridization between the $d_{xy}$ and $d_{x^2 - y^2}$ bands
forbidden in the tetragonal symmetry. In the distorted structure,
these two states are both at the now-equivalent $\Gamma$/X point
of the downfolded zone and repel each other. Furthermore,
rotations disrupt the Ru-O $pd\sigma$ hopping more strongly than
the $pd\pi$ hopping and thus the $d_{x^2 - y^2}$ band moves down
relative to the $d_{xy}$ band. Both effects together lead to the
formation of a strongly mixed state at the $\Gamma$/X point which
moves down rapidly and gains more $d_{x^2 - y^2}$ character with
rotation angle. While it is 300 meV above E$_F$ for $\theta$ =
0$^\circ$, it crosses E$_F$ for $\theta$ = 7$^\circ$, and is the
origin of $\Phi$. Although $\Phi$ was not observed at $\Gamma$,
this absence is not surprising when considering the unfavorable
photoemission matrix elements due to the significant $d_{x^2 -
y^2}$ and $d_{xy}$ character of this state.

The effects of FM on the surface electronic structure were
evaluated by performing constrained fixed spin moment calculations
for the 6$^{\circ}$ surface with an imposed magnetization of 1
$\mu_{B}$ / Ru, a value consistent with Ref.\onlinecite{matzdorf}.
The corresponding FM surface calculation is shown in Figure
\ref{bands}b and is radically different from what is measured
experimentally; for instance, both the evHs at M and the bottom of
the $d_{xy}$ / $d_{x^2 - y^2}$ band at X are absent. Regardless of
the particular details of the calculations, such as the position
of the chemical potential and the bands, even the number of bands
expected and measured is in disagreement, thus favoring the NM
scenario.

Although exact comparisons between the theoretical calculations
and the ARPES data can be somewhat difficult due to the
significant electron-electron interactions, the qualitative
comparison of the ARPES data with the general behavior of the
calculated electronic structure should be robust. The earlier
conclusion of surface FM \cite{matzdorf} was based on the
comparison of structural data ($\theta$ = 9$^\circ \pm 3^\circ$)
to magnetic band structure calculations. However, the error bars
in the structural data are comparable to the spread in the
calculated rotation angles for a NM (6.5$^\circ$), AFM
(6.5$^\circ$), and FM surface (9$^\circ$), leaving room open for
alternative interpretations of the data. Furthermore, the
generalized gradient approximation employed in the aforementioned
calculations may be inclined to overestimate the tendency towards
magnetism, and even incorrectly predicts ferromagnetism in bulk
Sr$_2$RuO$_4$. \cite{degroot} We can place a maximum upper bound
on the strength of any existing FM by considering our experimental
resolution and the width of the quasiparticle peaks. If we assume
that both $\alpha_S$ and $\alpha_{F}$ were comprised of a pair of
extremely weakly split FM bands, we are able to put an upper bound
of E$_{exch}$ $\approx$ 15 meV, which is much smaller than the
predicted FM exchange splitting of $\approx$ 500 meV.
\cite{degroot,fang} Using this value of E$_{exch}$ $\approx$ 15
meV results in an upper bound for the spin polarization of $<$
0.05 $\mu_B$ / Ru, much weaker than predicted theoretically.

In conclusion, we have isolated and directly studied the
surface-derived electronic states in Sr$_2$RuO$_4$ by ARPES. By
comparison with detailed band structure calculations, we find that
the origin of the ARPES features can be simply explained by
considering the effect of a nonmagnetic surface reconstruction on
the electronic structure, with no evidence of surface FM. Although
we conclude that the surface is nonmagnetic, we believe that our
finding does not necessarily disfavor FM pairing mechanisms, and
that both FM and AFM fluctuations most likely still exist in the
bulk.

We thank Z. Fang and K. Terakura for kindly discussing their
unpublished band structure calculations with us. The work at NRL
is supported by the ONR. SSRL is operated by the DOE Office of
Basic Energy Research, Division of Chemical Sciences. The office's
division of Material Science provided support for this research.
The work at Stanford University was also supported by NSF Grant
No. DMR9705210 and ONR Grant No. N00014-98-1-0195.


\begin{references}
\bibitem{velocityratio} $v_{F}^{B} / v_{F}^{S}$ is larger than calculated.
        This may be due to the relaxation of the
        apical oxygens, additional tilting of the RuO$_6$ octahedra,
        or even different electron-phonon coupling at the surface, but
        does not change our arguments.
\end{references}
\end{document}